\documentclass[usenatbib]{mnras}
\usepackage{graphicx}
\usepackage{txfonts}

\usepackage{url}
\usepackage{graphicx}
\usepackage[hyphenbreaks]{breakurl}
\usepackage{float}
\usepackage{subfig}
\usepackage{multirow}
\usepackage{natbib}

\title[Spicules Detection]{Detection of spicules termed Rapid Blue-shifted Excursions as seen in the chromosphere via H$\alpha$ and the transition region via Si {\sc iv} 1394 \AA\ line emission.}

\author[Vilangot Nhalil et al.]{Nived Vilangot Nhalil,$^{1,2}$\thanks{E-mail: Nived.Vilangot.Nhalil@armagh.ac.uk}
Juie Shetye$^{3,1}$ , J. Gerry Doyle$^{1}$ \\
$^{1}$Armagh Observatory \& Planetarium, College Hill, Armagh, BT61 9DG, N. Ireland\\
$^{2}$Astrophysics Research Centre (ARC), School of Mathematics and Physics, Queens University, Belfast, BT7 1NN, N. Ireland\\
$^3$Department of Astronomy, New Mexico State University, Las Cruces, 88001, USA }

\begin{document}
\outer\def\gtae {$\buildrel {\lower3pt\hbox{$>$}} \over 
{\lower2pt\hbox{$\sim$}} $}
\outer\def\ltae {$\buildrel {\lower3pt\hbox{$<$}} \over 
{\lower2pt\hbox{$\sim$}} $}
\newcommand{\Msun}{$M_{\odot}$}
\newcommand{\lsun}{$L_{\odot}$}
\newcommand{\Rsun}{$R_{\odot}$}
\newcommand{\solar}{${\odot}$}
\newcommand{\kep}{\sl Kepler}
\newcommand{\ktwo}{\sl K2}
\newcommand{\tess}{\sl TESS}
\newcommand{\swift}{\it Swift}
\newcommand{\Porb}{P_{\rm orb}}
\newcommand{\nuorb}{\nu_{\rm orb}}
\newcommand{\eplus}{\epsilon_+}
\newcommand{\eminus}{\epsilon_-}
\newcommand{\cd}{{\rm\ c\ d^{-1}}}
\newcommand{\MdotL}{\dot M_{\rm L1}}
\newcommand{\Mdot}{$\dot M$}
\newcommand{\Mdotsolar}{\dot{M_{\odot}} yr$^{-1}$}
\newcommand{\Ldisk}{L_{\rm disk}}
\newcommand{\src}{KIC 9202990}
\newcommand{\ergscm} {erg s$^{-1}$ cm$^{-2}$}
\newcommand{\rchi}{$\chi^{2}_{\nu}$}
\newcommand{\chisq}{$\chi^{2}$}
\newcommand{\pcmsq} {cm$^{-2}$}

\newcommand\Ion[2]{#1$\;${\scshape{\uppercase\expandafter{\romannumeral#2}}}}
\providecommand{\lum}{\ensuremath{{\cal L}}}
\providecommand{\mg}{\ensuremath{M_{\rm G}}}
\providecommand{\bcg}{\ensuremath{BC_{\rm G}}}
\providecommand{\mbolsun}{\ensuremath{M_{{\rm bol}{\odot}}}}
\providecommand{\teff}{\ensuremath{T_{\rm eff}}}

\maketitle
\begin{abstract}
We show signatures of spicules termed Rapid Blue-shifted Excursions (RBEs) in the Si {\sc iv} 1394 \AA\ emission line using a semi-automated detection approach. We use the H$\alpha$ filtergrams obtained by the CRISP imaging spectropolarimeter on the Swedish 1-m Solar Telescope and co-aligned Interface Region Imaging Spectrograph data using the SJI 1400 {\AA} channel to study the Spatio-temporal signature of the RBEs in the transition region. The detection of RBEs is carried out using an oriented coronal loop tracing algorithm on H$\alpha$ Dopplergrams at $\pm 35$ km s$^{-1}$. We find that the number of detected features is significantly impacted by the time-varying contrast values of the detection images, which are caused by the changes in the atmospheric seeing conditions. We detect 407 events with lifetime greater than 32 sec. This number is further reduced to 168 RBEs based on the H$ \alpha$ profile and the proximity of RBEs to the large scale flow. Of these 168 RBEs, 89 of them display a clear Spatio-temporal signature in the SJI 1400 {\AA} channel, indicating that a total of $\sim$53$\%$ are observed to have co--spatial signatures between the chromosphere and the transition region.
\end{abstract}


\section{Introduction}

Solar spicules are dynamic jets of plasma best seen in the Sun's chromosphere. They move upwards with velocities ranging from 15--200 km s$^{-1}$, lasting from tens of seconds to a few minutes. They are thought to play an important role in the mass and energy transfer across the solar atmosphere. Usually this transport is in the form of propagating magneto-hydro dynamics (MHD) waves \citet[]{Roberts_1945, DePontieu_2007a,Zaqarshvili_2008,McIntosh_2011,Shetye_2021} or at sites of magnetic reconnection, see \citet[][and references therein]{DePontieu_2007c,Langangen_2008,Samanta_2019}. In addition, spicules are further speculated to play a role in the formation of the solar wind \citep{bart_2007,McIntosh_2011} which interacts and can modulate Earth’s magnetic field. These magnetic structures are believed to be emerging from the interiors of granules, with their movement governed by photospheric motions \citep{DePontieu_2007c}. 

Quantifying the number density of spicules which play a role in mass and energy transfer across the solar atmosphere is important towards understanding their role in the energy budget. There has been multiple efforts to track these features across the solar atmosphere, e.g. \citet{Pereira_2014} showed that some spicules (termed Type-II or rapid blue-shifted excursions 
\citep[RBE][]{Langangen_2008,sekse_2013a,sekse_2013b} undergo thermal evolution, at least to transition region temperatures. \citet{Henriques_2016} suggested that at least 11$\%$ of RBEs could be connected to coronal counterparts. Work by \citet{srivastava_2017} showed that these features support high-frequency waves that can transfer mass and energy into the solar corona. 

Ideally spicules are observed in the chromosphere in the wings of H$\alpha$ and Ca {\sc ii} K spectral lines \citep{Tsiropoula_2012}. In the transition region, signatures of spicules often appear in Si {\sc iv} and C {\sc ii} slit-jaw images as bright elongated features \citep{Tian_2014}. However, the co-spatial origins of these bright elongated features with the lower atmospheric features, still remains debated. In this manuscript we revisit the on-disk counterparts of spicules, such as the RBEs, where we improve our detection strategies using the Oriented  Coronal Curved Loop Tracing (OCCULT) \citep{aschwanden_2013, aschwanden_2010} code. We play special attention to the shape of the events, their origin from the photosphere and their spectral profiles. We use data from the Swedish 1-m Solar Telescope \citet[][SST]{Scharmer_2003} and Interface Region Imaging Spectrograph (IRIS) whose field-of-view (FOV) was focused below a pore within active region NOAA AR 12080. In particular we look at the best method to investigate the evolution from the 
chromosphere as observed using H$\alpha$, to the transition region as seen in Si {\sc iv} 1394 \AA. 
\begin{figure*}
\vspace*{-1.5cm}
\centering
\includegraphics[width=0.99\textwidth]{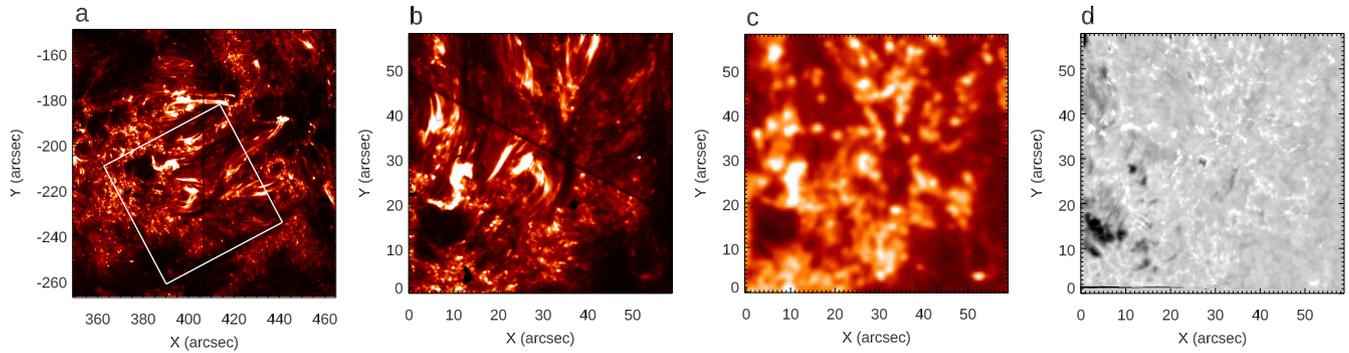}
\vspace{-0.8cm}
\caption{Panel (a): SJI 1400 \AA\ image of the region under study. The white box bounds the SST FOV. Panel (b): The co-aligned SJI FOV. Panel (c) The 
AIA 1600 \AA\ image. Panel (d) The SST image obtained at --1.032 \AA\, from the line centre. } 
\label{fig1}
\end{figure*}
This follows on from work by \cite{Nived2022} who addressed the issue as to why there is not always a one-to-one correspondence, between Type II spicules and hot coronal plasma signatures. These authors do not detect any difference (as seen in H$\alpha$), in their spectral properties in a quiet Sun region compared to a region dominated by coronal loops. Although the number density close to the foot-points in the active region is found to be an order of magnitude higher than in the quiet Sun case. 
\section{Observations and detection methodology} \label{ob}
\subsection{Data sources}

We used coordinated observations from SST and IRIS, with support from the Solar Dynamics Observatory (SDO) to track and study the evolution of RBE. The ground-based observations analysed in this article were obtained on 10$^{th}$ June 2014, in both an imaging and a spectroscopic mode, using the CRISP \citep{Scharmer_2008} instrument on the Swedish 1-m Solar Telescope \cite[][SST]{Scharmer_2003}. CRISP has a FOV of 60$''$ $\times$ 60$''$ with an approximate pixel scale of 0.0592$''$. After co-alignment with SDO/AIA 170.0 nm images, we determined that the CRISP FOV is tilted at an angle of 62$^{\circ}$04$'$ with respect to the SDO heliocentric FOV, and centered at $xc = 403''$ and $yc = -211''$ which contained a pore and was entirely within NOAA AR 12080. Nine H$\alpha$ line-positions were sampled in sequence up to $\pm$1032 m\AA\ with equidistant intervals of 258 m\AA\ from the line core at 6562.8\AA\ at position 0. The cadence of the observations is $\sim$4 s. More details about the observations can be found in \cite{shetye_2016}, including the Multi-Object Multi-Frame Blind Deconvolution \cite[MOMFBD]{Noort_2005} to produce science-ready images.

\begin{figure}
\centering
\includegraphics[width=0.5\textwidth]{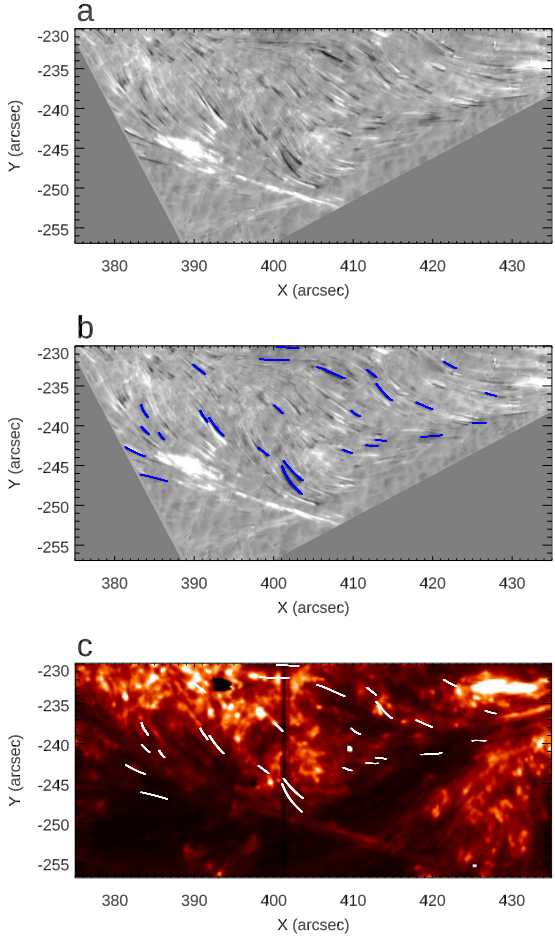}
\caption{Here we present the tracing results for one SST image observed at time 07:57.12 UT. Panel (a): The Dopplergram obtained by subtracting the red wing from the blue wing. The black pixels represents the location of the RBEs. Panel (b): Same as panel (a). The blue curves indicate the axis of the RBEs detected from the curve-linear tracing algorithm. Panel (c): The same FOV in SJI 1400 {\AA} channel with detected features overlaid in white colour.} 
\vspace{1cm}
\label{fig2}
\end{figure}

IRIS observed active region NOAA 12080 in a sit-and-stare mode and obtained Si~{\sc iv} 1394~\AA\ spectra using the $0.33 "$ wide slit. The exposure time of the observation is $\sim$15 s, making a cadence of $\sim$16 s. The pixel scale of IRIS is $0.16635''$, which is approximately 3 times the SST pixel scale. IRIS cadence is, therefore, $\sim$4 times that of SST, thus, there are four SST observations corresponding to one IRIS observation. A near-perfect co-alignment between IRIS and SST must be achieved to study the Si~{\sc iv} signature corresponding to spicules as seen in H$\alpha$. The co-alignment is performed using the AIA 1600~\AA\ and SJI 1400~\AA\ channels via a cross-correlation technique. The results of the SST-IRIS co-alignment are presented in Figure \ref{fig1}. Panel (a) shows the FOV of the SJI 1400 \AA\ observations, while the white box bounds the FOV of the SST observation. The co-aligned images of the SJI 1400 \AA\ and AIA 1600{\AA} are shown in panels (b) and (c) respectively. Panel (d), shows the H$\alpha$ image at --1.032 {\AA} from the line centre.

The detection of RBEs is carried out using H$\alpha$ Dopplergrams. This is a commonly used technique to detect RBEs and RREs \citep{sekse_2013}. The Dopplergram is constructed by subtracting opposite H$\alpha$ wing position at line positions $\pm$773.9~m\AA\ (i.e. $\sim \pm$35 km s$^{-1}$). Before the subtraction, the line intensities at these wavelength positions are normalised with respect to the outermost wavelength observed in H$\alpha$ 
at $\pm$ 1.032 {\AA} from the line centre. An example of a Dopplergram is shown in panel (a) of Figure \ref{fig2}. Since we subtract the red wing from the blue wing, RBEs appears as a dark region in the Dopplergram, while the RREs appear as a bright region. The next task is to obtain properties such as lifetime, and length of these events from the Dopplergrams. For this, we need to trace the structure axis of the RBEs. This is done via a curvi-linear structure tracing algorithm called Oriented  Coronal Curved Loop Tracing (OCCULT) \citep{aschwanden_2013, aschwanden_2010}. The code is routinely used for detecting loop structures in EUV images obtained from instruments such as TRACE and AIA. \cite{aschwanden_2013} has shown that the automated loop tracing code can also be applied on H$\alpha$ line core image data to trace filamentary structures and spicules. In this work, we follow a similar analysis to trace RBEs from H$\alpha$ Dopplergrams. The details of the tracing algorithm are explained below.
 

\subsection{Visual and statistical analysis using the H$\alpha$ line}{\label{detection}}

Here we outline the detection scheme. The code starts by suppressing the background structures below a certain threshold. The threshold ($I_{thresh-1}$) is given by,
 \begin{equation}
I_{thresh-1}=I_{med}\times q_{med}
\end{equation}
where $I_{med}$ is the median of the intensity image and $q_{med}$ is a control parameter. The value of $q_{med}$ is normally between 1 to 2.5, depending on the noise level in the data \citep{aschwanden_2013}. This step is useful for removing faint background features which appears due to the noise. Compared to the typical EUV images, the CRISP images have a very high contrast, implying that they have a broad intensity distribution. Therefore, we chose $q_{med} =0$ which means that every pixel will be considered for the curve linear detection. The code has an option to perform low-pass and high-pass filtering on the detection image before the tracing begins. Filtering is useful in reducing noise and enhancing the faint structures. Our goal is to trace the dark and bright features in the Dopplergram. The dark features appear with an intensity reduction of $153 \%$ compared to the background. Therefore, these features are easily detected without any filtering.

Tracing of the feature starts from the darkest pixel in the image. Once a structure is traced, the location of the structure will be set to zero before the next tracing begins. This iterative procedure continues until there are no pixels above a certain intensity threshold ($I_{thresh-2}$) where $I_{thresh-2}$ is given by,
 \begin{equation}
 \label{eqn_2}
I_{thresh-2}=I_{med}\times q_{thresh-2}
\end{equation}
where $q_{thresh-2}$ is another control parameter. We set $q_{thresh-2} = 3$ for the tracing of RBEs from the Dopplergrams (this value was determined via a trial-and-error.)

The OCCULT code traces separately in both the forward and backward direction from the initial starting point, then finally combined to form the complete structure. Once the starting point is identified, the code searches the direction ($\theta$, with respect to the x-axis), where the flux is maximum and then obtains the local curvature radius ($r_{m}$) corresponding to the maximum flux. Following this, the code uses a second order guiding criterion to extrapolate the coordinates of the traced segments; for a detailed mathematical description of the code, see \cite{aschwanden_2013}. 
Note that the slit can be placed at any angle between 0 to 180 with respect to the x-axis, however only one angle corresponds to the local direction of the feature. This local direction can be obtained by calculating the average flux along the slit for each orientation of the slit. The tracing stops when the pixel below $I_{thresh-2}$ is reached. 

The RBEs traced using the OCCULT code are presented in panel (b) of Figure \ref{fig2}. The code successfully traced several RBEs from the Dopplergram and their coordinates are shown in blue. To avoid a false detection, we applied a condition that the length of the traced structure should be greater than 20 pixels (1.18") and they must have a minimum radius of curvature of 40 pixels (2.36"). From the output of the OCCULT code, we have information about the location of RBEs in each time frame. By following an iterative approach it is possible to identify the starting and end time of RBEs.

\begin{figure}
\centering
\includegraphics[width=0.5\textwidth]{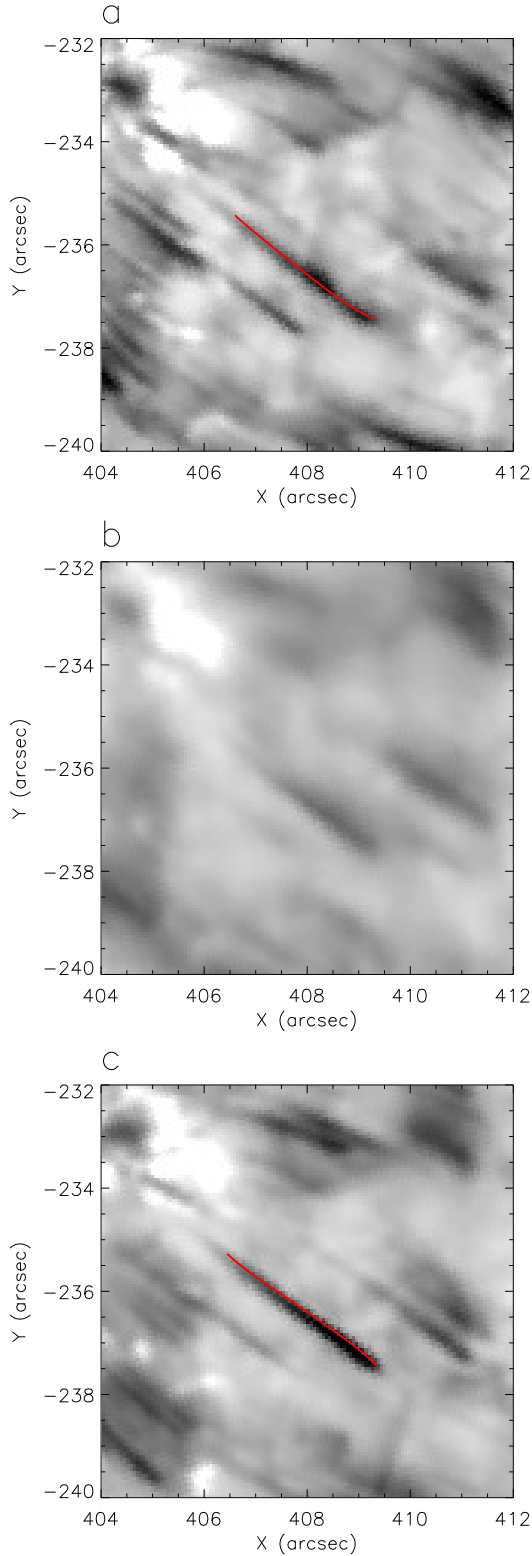}
\caption{The effects of blurriness on RBE detection. Panel (a): The red curve shows the axis of RBE detected using curve-linear tracing algorithm. Panel (b): The same region as in panel (a) after 36 s. Notice that the detection algorithm fails detect the RBE due to the blurriness caused by bad seeing condition. Panel (c), The same region as in panel (b) after 84 s.} 
\label{fig3}
\end{figure}

\begin{figure}
\centering
\includegraphics[width=0.5\textwidth]{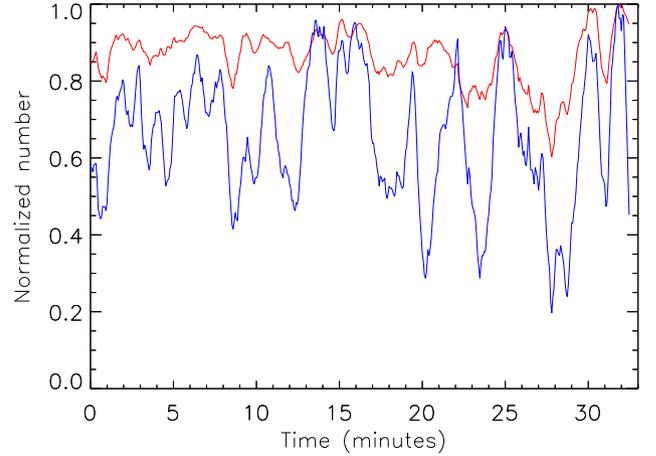}
\caption{The red curve represents the time evolution of the contrast value. The blue curve shows the number RBEs detected by the curve-linear tracing algorithm as a function of time. Both curves are well correlated, indicating the effects of contrast value on the detection algorithm.}
\label{fig4}
\end{figure}

\begin{figure*}
\centering
\includegraphics[width=\textwidth]{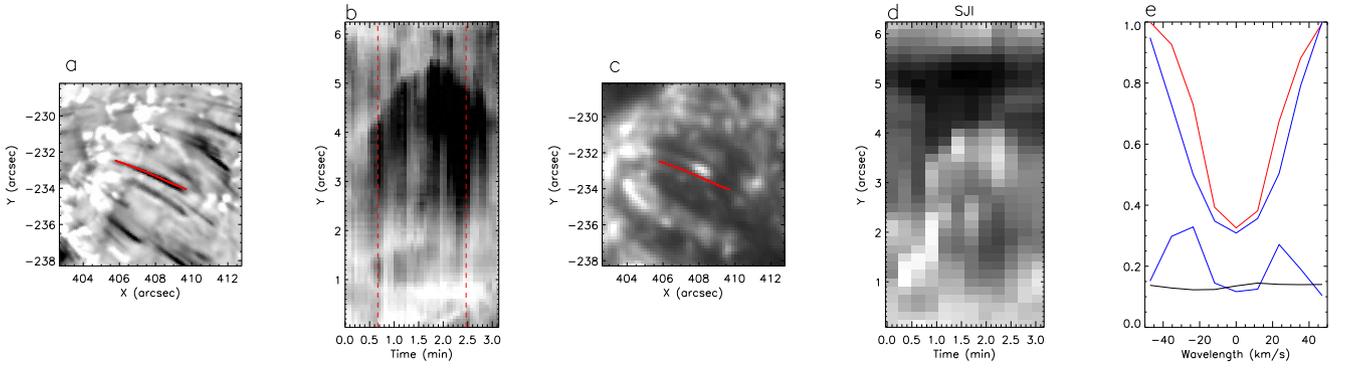}
\caption{The comparison of SJI and H$\alpha$ time-distance map. Panel (a): H$\alpha$ blue wing image (--35 km s$^{-1}$) at the time frame in which the RBE appeared maximum in its length. The red curve marks the axis of the RBE obtained from the detection algorithm. Panel (b): The time-distance map of the H$\alpha$ blue wing image (--35 km s$^{-1}$). The red dotted line indicates the starting and ending time of the RBE obtained from the detection algorithm. The same region as in panel (a) in the SJI 1400 {\AA} channel. The red curve shows the the axis of the RBE. Panel (d): Time-distance map in the SJI 1400 {\AA} channel. Panel (e): The averaged profile of the RBE (in blue) derived from the pixels through which the red curve passes in panel (a). The red profile corresponds to the quiet Sun. The blue curve at bottom indicates the absolute residual profile, the 3 sigma level is marked in black colour.  } 
\vspace{1cm}
\label{td1}
\end{figure*}

\begin{figure*}
\centering
\includegraphics[width=\textwidth]{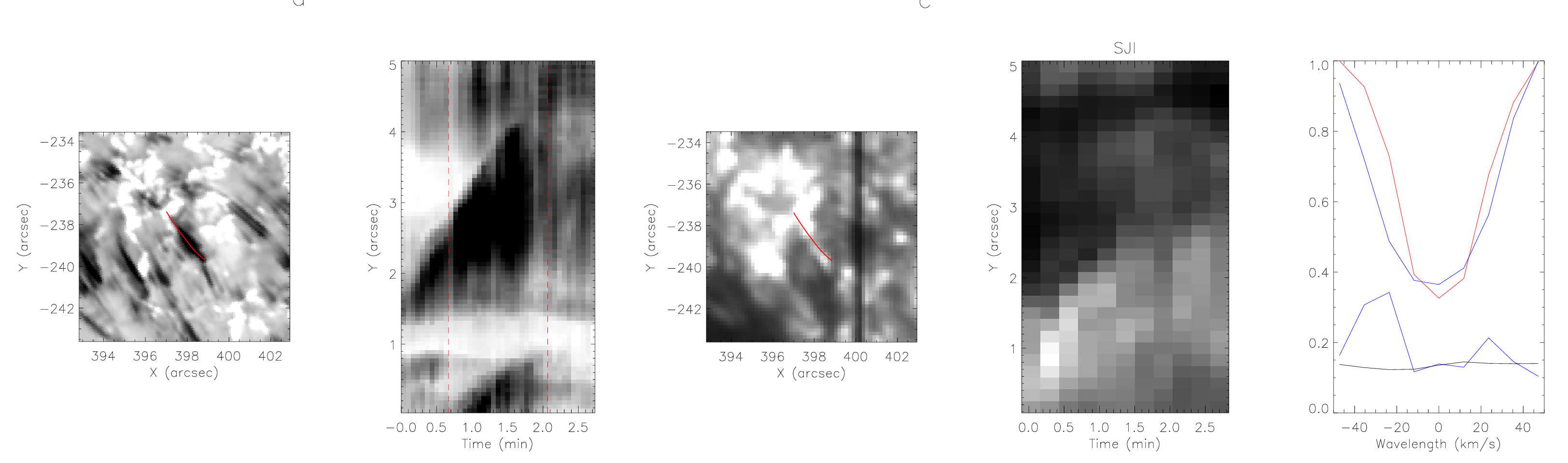}
\caption{Same as Fig~\ref{td1}, but for another event.} 
\vspace{1cm}
\label{td2}
\end{figure*}
We start by identifying the location of an RBE in the first time frame (i.e. time frame 1), then in the second step, it goes to the next time frame (i.e time frame 2), finding all RBEs within a 200 km range. Then identify the RBE that is closest. If the average minimum distance between the RBEs in the first frame and the one in the next frame is less than 5 pixels ($\sim0.3"$) then it will be considered the continuation of RBE in the previous time step. The average minimum distance is the average of the minimum distance between the coordinates of the RBEs in the previous frame to the one in the next frame. This iteration continues until the RBE completely disappears (or RBE intensity goes below $I_{thresh-2}$, see equation \ref{eqn_2}). However, the RBEs could disappear due to the bad quality (low contrast) of the image. The main cause of this blurriness is the time varying atmospheric seeing condition.
A good seeing condition is characterized by higher contrast values of the observed images, while the low contrast values corresponds to the bad seeing condition.
To account for the effects seeing on the time-tracing algorithm, the iteration only stops when the RBEs disappears on a high contrast image. Here, we 
defined contrast value as the ratio of the standard deviation to the intensity of the image. By manual inspection, we defined the high contrast image 
as the image with a contrast value greater than 1/18. Once an RBE is traced in time, then its coordinates will be removed from the data cube to avoid 
multiple detection. A detailed description of the effects of contrast values on the detection algorithm is presented in the next section. Finally, we 
applied a condition that the lifetime of an RBEs should be at least 2 times-steps ($\sim$8 s), to reduce the number of false detections.


\subsection{The effects of blurriness on the detection}

The detection algorithm mentioned in the previous section significantly depends on the quality (contrast) of the image used for detecting RBEs.  The contrast value will be smaller for the blurred image. This blurriness could be a result of atmospheric seeing, movement of the instrument or even related to the data-reduction pipeline. 

An example of the effect of blurriness on the detection algorithm is presented in Figure \ref{fig3}. Panel (a) shows an RBE observed on a good contrast image. As mentioned before, good contrast images are images with a contrast value greater than 1/18. The red curve indicates the axis of the RBE obtained from the curve-linear tracing algorithm. Panel (b), shows the same region as in panel (a), but after 36 s. The contrast value for this image is $\sim{1/27}$. The image appears completely blurred and the detection algorithm fails to detect the RBEs in this frame. When the contrast value increases again, the RBEs reappear and are detectable by the detection algorithm as shown in panel (c). One must account for this effect when calculating the lifetime of the event. Otherwise, it will lead to an erroneous estimate of lifetime and multiple detections of the same RBE.

The dependence of the number of detected RBE on the contrast value is evident in the time evolution plot shown in Figure \ref{fig4}. The red curve indicates the time evolution of the contrast value. The curve was smoothed by a window of 8 frames (32 s) in length, to bring out the background variation. The lower value corresponds to the blurred images and a period of blurriness appears as a trough in the time evolution of the contrast value. The blue curve depicts the time evolution of the number of RBEs detected by the curve-linear tracing algorithm. Similar to the red curve, the blue curve is normalised as well as smoothed by a window 8 frames in length. A period of less number of detection matches very well with a period blurriness. Both the curves are well correlated with a Pearson correlation coefficient of 0.75. This confirms the role of blurriness on the detection algorithm.

\section{Methods}
The main goal of this paper is to identify the signature of spicules in Si~{\sc iv} 1394 \AA. To do this we first created a movie based on data from the SJI 1400 {\AA} channel, over-ploting the axis of the RBEs obtained from the detection code described in Section \ref{detection}. The movie reveals that brightenings in the SJI 1400 {\AA} channel form in the proximity of spicules and some of these brightenings appears to evolve along the axis of the spicule. We quantify Si~{\sc iv} signatures associated with spicules by following two different methods.

\subsection{Preliminary identification method}
The first method is based on the fraction of spicules that shows a nearby SJI 1400 {\AA} brightening. A brightening in the 1400 {\AA} channel is defined as a 3$\sigma$ intensity enhancement above the background level as described in \cite{paper_nano_flare}. If a brightening appears within the 5-pixel (0.8$"$) location of the RBE at any point in its lifetime, then we assume that the brightening is associated with the RBE. For this analysis, we excluded all the single frame RBEs to reduce the number of false detections. There were 1936 detections with a lifetime of at least 2 time frames. Out of the 1936 RBEs; 639 ($\sim33\%$) of them were within the 5-pixel location of the SJI brightening. This fraction should be interpreted with caution due to the following reasons. 

\begin{figure*}
\centering
\includegraphics[width=\textwidth]{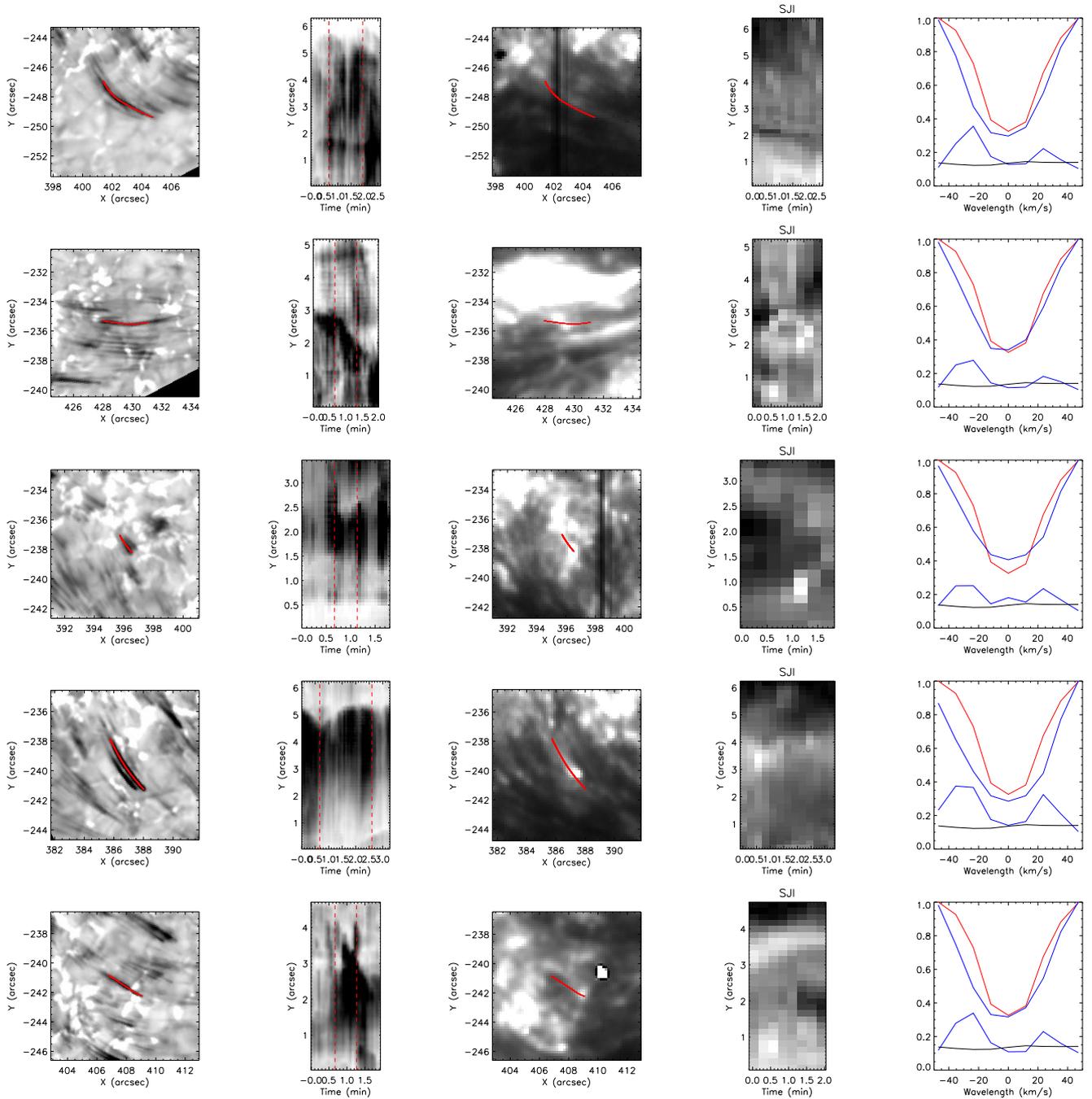}
\caption{The time evolution of different categories of RBEs presented in the pi diagram (Fig. \ref{pi}). The Figure format is the same as in Fig.  \ref{td1}. 1st row: RBEs in the proximity of the flow. 2nd row: an RBE like event close to a flaring region. We exclude this type of events because the 
time evolution in the SJI 1400 {\AA} channel is dominated by the flaring region. 3rd row: RBEs with significant enhancement in the line core intensity. 
4th row: RBEs with stronger absorption in the core. 5th row: RBEs without a spatio-temporal signature in the SJI 1400 {\AA} channel.} 
\vspace{1cm}
\label{td3}
\end{figure*}

The SJI 1400 {\AA} image of active region plage is abundant with small scale brightenings. Therefore it is difficult to conclude that all of the intensity enhancements close to the RBEs are associated with the RBEs. For example, we notice that some of the brightenings appear even before the formation of RBEs and its lifetime has no clear correspondence with the lifetime of the RBEs. Furthermore, the formation of multiple RBEs within the 5-pixel location of the brightening may lead to an over-estimation of the above-calculated fraction. Therefore, a method that incorporates the time evolution of RBEs is essential to study its corresponding signature in SJI 1400 {\AA}. This is described in the following section.
\begin{figure*}

\centering
\includegraphics[width=0.9\textwidth]{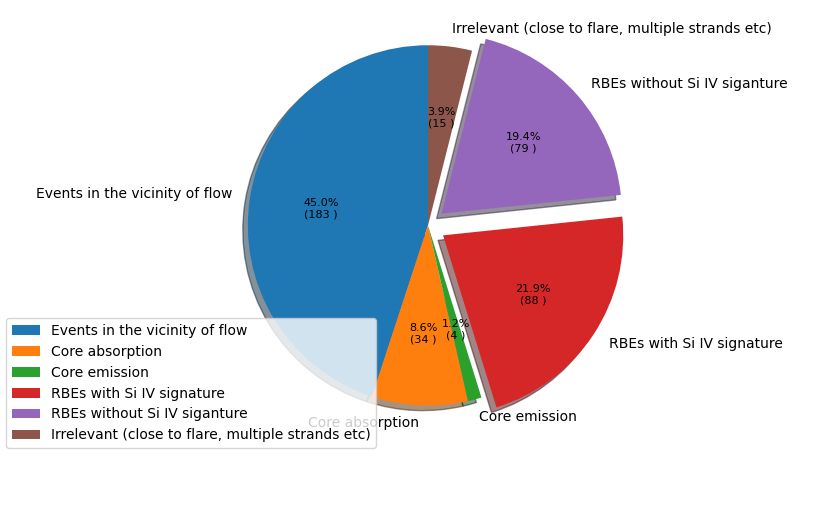}
\vspace*{-1cm}

\caption{Rapid Blue-shifted Excursions statistics in H$\alpha$ and Si {\sc iv} 1394 {\AA} line emission.} \label{pi}
\end{figure*}

\subsection{Refined identification method}
Refined identification for quantifying the Si~{\sc iv} signature of RBEs incorporates both the time and spatial evolution of the RBE and the corresponding Si~{\sc iv} signature. If the spatio-temporal evolution matches, then we can assume that the Si~{\sc iv} signature is indeed the transition region counterpart of the RBEs. To do this, we constructed the time-distance plot of the H$\alpha$ blue wing and SJI 1400 {\AA} by placing a 1.6$"$ wide slit along the axis of the RBE. A time-distance plot can provide information about the simultaneous evolution of spicules in SJI 1400 \AA\ and the H$\alpha$ blue wing. By comparing these two plots we can detect signatures of spicules in the SJI 1400 {\AA} channel. The cadence of the SJI observation is 16 s, therefore to see a Si~{\sc iv} signature in a time-distance plot, the RBEs should have a minimum lifetime of 2 SJI time frames (32 s). Due to this limitation of cadence, only RBEs with a lifetime greater than 32 s were chosen for this analysis. Only 407 detections satisfy the above lifetime condition.
 
We then created the time-distance plot for those RBEs by placing a $1.6"$ wide slit along its axis. The H$\alpha$ blue wing time-distance plot was 
constructed by plotting the minimum intensity along the width of the slit at each length. In the SJI 1400 {\AA} channel, we used the maximum intensity.
Then, we studied the similarity between H$\alpha$ and SJI 1400 {\AA} time-distance plot by manual inspection. Out of the 407 detections, almost half of these events (183) were close to a large scale active region flow. The events observed at this location appeared very similar to an RBE profile. However, their shape was irregular and showed a large scale motion. The signature of a large scale mass motion was also visible in the SJI 1400 {\AA} channel. Therefore, we excluded those events from the rest of the analysis. Then we checked the H$\alpha$ profile of the remaining events. The profile was obtained by averaging the H$\alpha$ profiles along the axis of the RBEs at the time frame in which the RBE appeared maximum in its length. Some of these events (35) had a significant core absorption when compared with a quiet Sun profile while another 5 showed enhancement in line core intensities compared to the QS profile. Here, absorption and emissions mean the intensities are lower and higher than the background intensities respectively. These profiles cannot be considered as an RBE profile. Hence, we removed those events with absolute residual value above 3$\sigma$ at the line core. The residual curve was calculated by subtracting the H$\alpha$ profile from the quiet Sun profile. There were 168 RBEs, that satisfies all the above conditions. The comparison of time-distance plot was performed on these remaining 168 RBEs.

\section{Results and Discussion} 

Two examples of time-distance plots are presented in Figures \ref{td1} and \ref{td2}. Panel (a) shows the location of the spicule in the H$\alpha$ blue wing, while panel (b) shows the corresponding time-distance map. The time frame chosen for the H$\alpha $ image in panel (a) corresponds to the frame in which the RBE appeared maximum in its length. The SJI 1400 {\AA} channel counterpart of the spicule is shown in panel (c) and its time-distance map is presented in panel (d). The red curve in panels (a) and (c) indicates the axis of the RBE obtained from the detection code. A bright jet-like feature can be seen at the location of the RBE in the SJI 1400 {\AA} channel. Therefore, in the SJI 1400 {\AA} channel, we use the maximum intensity along the width of the slit to create the time-distance plot, while in the H$\alpha$ blue wing, we use the minimum intensity. The length of the slit was chosen based on the maximum length of the spicule and is equal to the maximum length of the RBEs, i.e +1$"$ 
Finally, in panel (e), we plot the averaged H$\alpha$ profile (blue) obtained from the pixels along the axis of the RBEs as shown in panel (a). A quiet Sun profile is plotted in red. The blue curve at the bottom shows the absolute residual profile obtained by subtracting the averaged H$\alpha$ profile from the quiet Sun profile. The black curve indicates the 3$\sigma$ level.

In Figure \ref{td1}, the time-distance path of the RBE follows a clear parabolic shape as seen in both the SJI 1400 {\AA} channel and in the H$\alpha$ blue wing, indicating that the jet-like shape seen in the SJI 1400 {\AA} channel is the transition region counterpart of the RBE. The time-distance path seen here is similar to the trajectories of the dynamic fibrils seen in the H$\alpha$ line core as reported by \cite{Hansteen_2006} and \cite{DePontieu_2007a}. It is suggested that the parabolic trajectories of the dynamic fibril are caused by shock waves which are driven by the leakage of magneto-acoustic oscillation along the inclined field lines \citep{Hansteen_2006, DePontieu_2007a, Heggland_2007}. However, recent studies by \cite{Pereira_2014} and \cite{skogsrud_2015} suggest that the Type II spicule with weak (incomplete) parabolic trajectories in Ca {\sc ii} H and a clear parabolic path in AIA and SJI passbands may not be driven by shock waves. Instead, they could be generated by a magnetic reconnection process. In Figure \ref{td1}, the time-distance plot in the SJI 1400 {\AA} channel also shows the presence of a sudden vertical brightening as seen in \cite{De_pontieu_2017a}. With the aid of 2.5 D MHD simulations, \cite{De_pontieu_2017a} demonstrate that these rapid apparent motions are the signatures of the heating front. On the other hand RBEs in Figure \ref{td2} disappear suddenly from the H$\alpha$ passband. However, a weak presences of down-falling material can still be traced in H $\alpha$ time-distance plot, which makes it appears as an incomplete parabola in the time-distance plot. A similar parabolic shape can also be traced in the time-distance plot of the SJI 1400 {\AA} channel, indicating that this RBE has a transition region counterpart. The transition region counterpart continues to evolve even after the RBE partially disappear from H$\alpha$. This is the classic behaviour of a type II spicule as shown in numerous off limb studies \citep{Pereira_2014,skogsrud_2015}.

The main difference with the refined method is that we now identify bright jets due to the Si~{\sc iv} line at the locations of some RBEs. These jets 
have the same morphology, in terms of shape and size as the RBEs. The time-distance plots, show the direction of motion corresponding to the RBEs, and the RBEs seems to have motion along the same direction. For many RBEs, we see a reversal in the direction at both heights. The motion of these RBEs across different heights have the similar slope. In some cases, there is a slight misalignment between the RBE and its signature in Si {\sc iv}. This is 
because the H$\alpha$ and SJI 1400 {\AA} channel sample different temperature regimes of the RBEs and thereby a different height in the solar atmosphere. Thus the signature in passbands that sample higher temperatures may not spatially coincide along the path of the RBE. 

Figure \ref{td3} presents an overview of different categories of RBEs observed in the active region plage. The columns are arranged same way as in Figure \ref{td1} and \ref{td2}. See the above figure for the description of each column. The top row shows the time-evolution of an RBE in the vicinity of 
the flow. A large scale flow is also visible in the SJI 1400 channel, thus the time-evolution in the SJI 1400 is dominated by the flow and cannot be considered for studying an RBE signature. Similarly, the RBE shown in the 2nd row is also excluded due to its proximity to the flaring region in the 
SJI 1400 {\AA} channel. The third row shows a rare case of an RBE with significant enhancement in line core intensities compared to the QS reference 
profile. While, the fourth row shows the time evolution of an RBE with significant absorption in the line core. The red wing is also significantly 
broadened for this event. However, the broadening in the blue wing is stronger than the red wing. This event has a co-spatial and co-temporal signature 
in the SJI 1400 {\AA} channel. The fifth row shows an example of an RBE with no co-evolving signature in Si {\sc iv}.

Final results on the statistical signature of RBEs in the transition region is summarised in Figure \ref{pi}. The statistical study was performed using 407 RBEs that have a lifetime of at least 2 SJI time frames (32s). Many (58$\%$) of these events were not considered when looking for the Si {\sc iv} signature due to the proximity of RBEs to the flow (45$\%$) and the enhanced absorption (8.6$\%$) and emission (1.2$\%$) of the RBEs at the H$\alpha$ line core. In the remaining 168 RBEs, 89 of them show a similar time-distance plot in both H$\alpha$ and SJI 1400 {\AA} channel. This implies that $\sim 53 \%$ of the selected RBEs have a clear counterpart in the transition region. This fraction is slightly lower compared to the study done by \cite{skogsrud_2015} on the off limb dataset observed by Hinode. \cite{skogsrud_2015} used a similar time evolution approach to study the signature of the off-limb spicule in the transition region. In their dataset 1, 71$\%$ of the Ca {\sc ii} spicules had a clear signature of Si {\sc iv} emission. Moreover, they find that 66 \% of the spicules have clear component in all four passbands (Ca {\sc ii}, Mg {\sc ii}, Si {\sc iv}, He {\sc ii}). \cite{Henriques_2016} studied the signature of spicules in the lower corona using on-disk H$\alpha$ observation in a quiet Sun region. They find that 11\% of the brightening detected in He {\sc ii} are associated with either RBEs or RREs. Furthermore, 6\% of the AIA 171 brightenings are linked to RBEs and RREs.

\section{Conclusions} \label{dis}
We study the statistical properties of RBEs and their signature in the transition region using co-spatial and co-temporal observations from CRISP/SST and IRIS. RBEs are the disk counterpart of the Type-II spicules \citep{Langangen_2008}. They are faster and short-lived compared to the traditionally observed Type-I spicules \citep{2007PASJ...59S.655D}. In H$\alpha$ and Ca {\sc II} observations, they appear with enhanced absorption in the blue wing compared to the red wing. Recent studies \citep{skogsrud_2015, Pereira_2014} using Hinode Ca II 3968.5 {\AA} filtergrams suggest that the Type-II spicule continue their evolution in hotter IRIS and AIA channels even after they seem to disappear from the cooler chromospheric passbands. This indicates that Type-II spicules undergo heating at least to transition region temperatures.
 
The RBE detection was carried out on H$\alpha$ Dopplergrams using the Coronal Curved Loop Tracing (OCCULT) routine \citep{aschwanden_2013}. We use the SJI 1400 {\AA} channel to investigate the co-temporal and co-spatial signature of the RBEs in the transition region of an active region plage. We note that the SJI 1400 {\AA} channel has a bandwidth of 40 {\AA}, therefore there could be contributions from the UV continuum, however, in the magnetic region such as an active region plage, the SJI 1400 {\AA} channel is dominated by the contribution from the resonance lines of Si {\sc iv} at 1394 and 1403 {\AA} \citep{skogsrud_2015}, which are formed at transition region temperatures of $\sim$ 85000 K.

In general, the feature detection algorithm depends on the contrast of the image in which the detection is carried out. We find a similar result by studying the correlation between the temporal changes in the contrast values and the number of detected events. This effect is quite common in any feature detection using ground based observations. Recently, \cite{Joshi_2022} found similar effects when detecting very small-scale quiet-Sun Ellerman Bombs. The temporal variation of contrast value is mainly caused by the time-varying atmospheric seeing condition. The algorithm fails to detect RBEs in time frames which are affected by bad seeing conditions. This effect was taken into account when calculating the lifetime of RBEs.

The detection algorithm obtained 1936 RBEs with a lifetime of at least 2 SST time-frames (8s). Almost 33\% of these RBEs were within the 5 pixel 
location of a 3$\sigma$ brightening in the SJI 1400 {\AA} channel. However, some of the SJI 1400 {\AA} brightenings appeared even before the formation of an RBE, furthermore the time evolution was independent of the lifetime of RBEs. This could mean that the observed brightness is not associated with an 
RBE and may lead to the erroneous estimate of the fraction. Therefore, we developed a refined method that includes the time evolution of both the 
Si {\sc iv} signature and the RBE to detect the signature of an RBE in the transition region. Based on this method, we find that 53\% of the selected RBEs have a co-temporal and co-spatial signatures in the chromosphere and the transition region. We detect 88 RBEs with a co-evolving signature in 
the Si {\sc iv} line during 32 minutes of observation. The area used for the RBE detection is 588 Mm$^2$. Using the lifetime of RBEs as 80s, we find that there are $3.8\times10^4$ RBEs with spatio-temporal signature in Si {\sc iv} at a given time on the solar surface. Considering the average lifetime of 
RBEs as 35 s, there are $3.6\times10^5$ RBEs on the solar surface at a given time. The present data does not allow to determine if the RBEs detected can 
serve as mediums for energy transport as we have not observed any particular wave-signatures associated with the events discussed in the paper. However complex wave-signatures associated with similar events are discussed in \cite{shetye_2016, srivastava_2017,Shetye_2021}.

In a followup paper we discuss spectral analysis related to these RBEs, studying details of the morphology and evolution of the co-spatial signatures across the solar atmosphere, and analysing the movement of plasma travelling along these jets between the chromosphere and the transition region.

\section*{Acknowledgments}
The authors are most grateful to the staff of the SST for their invaluable support with the observations. The Swedish 1-m Solar Telescope is operated on the island of La Palma by the Institute for Solar Physics of the Royal Swedish Academy of Sciences in the Spanish Observatorio del Roque de los Muchachos of the Instituto de Astrofísica de Canarias. Armagh Observatory \& Planetarium and NVN studentship is core funded by the N. Ireland Executive through the Dept. for Communities. The authors wish to acknowledge the DJEI/DES/SFI/HEA Irish Centre for High-End Computing (ICHEC) for the provision of computing facilities and support. The authors wish to acknowledge Markus Aschwanden for the use of OCCULT code.
We also like to thank STFC and the Solarnet project which is supported by the European Commission’s FP7 Capacities Programme under Grant Agreement number 312495 for T\&S. JGD would like to thank the Leverhulme Trust for a Emeritus Fellowship. JS is funded by NSF grant no: 1936336.

\section*{Data availability}
The reduced data underlying this article will be shared on reasonable request to the corresponding author.


\bibliographystyle{mnras}
\bibliography{ref}

\begin{thebibliography}{}
\makeatletter
\relax
\def\mn@urlcharsother{\let\do\@makeother \do\$\do\&\do\#\do\^\do\_\do\%\do\~}
\def\mn@doi{\begingroup\mn@urlcharsother \@ifnextchar [ {\mn@doi@}
  {\mn@doi@[]}}
\def\mn@doi@[#1]#2{\def\@tempa{#1}\ifx\@tempa\@empty \href
  {http://dx.doi.org/#2} {doi:#2}\else \href {http://dx.doi.org/#2} {#1}\fi
  \endgroup}
\def\mn@eprint#1#2{\mn@eprint@#1:#2::\@nil}
\def\mn@eprint@arXiv#1{\href {http://arxiv.org/abs/#1} {{\tt arXiv:#1}}}
\def\mn@eprint@dblp#1{\href {http://dblp.uni-trier.de/rec/bibtex/#1.xml}
  {dblp:#1}}
\def\mn@eprint@#1:#2:#3:#4\@nil{\def\@tempa {#1}\def\@tempb {#2}\def\@tempc
  {#3}\ifx \@tempc \@empty \let \@tempc \@tempb \let \@tempb \@tempa \fi \ifx
  \@tempb \@empty \def\@tempb {arXiv}\fi \@ifundefined
  {mn@eprint@\@tempb}{\@tempb:\@tempc}{\expandafter \expandafter \csname
  mn@eprint@\@tempb\endcsname \expandafter{\@tempc}}}

\bibitem[\protect\citeauthoryear{{Aschwanden}}{{Aschwanden}}{2010}]{aschwanden_2010}
{Aschwanden} M.~J.,  2010, \mn@doi [\solphys] {10.1007/s11207-010-9531-6},
  \href {https://ui.adsabs.harvard.edu/abs/2010SoPh..262..399A} {262, 399}

\bibitem[\protect\citeauthoryear{{Aschwanden}, {De Pontieu}  \&
  {Katrukha}}{{Aschwanden} et~al.}{2013}]{aschwanden_2013}
{Aschwanden} M.,  {De Pontieu} B.,   {Katrukha} E.,  2013, \mn@doi [Entropy]
  {10.3390/e15083007}, \href
  {https://ui.adsabs.harvard.edu/abs/2013Entrp..15.3007A} {15, 3007}

\bibitem[\protect\citeauthoryear{{De Pontieu} et~al.,}{{De Pontieu}
  et~al.}{2007a}]{DePontieu_2007c}
{De Pontieu} B.,  et~al., 2007a, \mn@doi [Science] {10.1126/science.1151747},
  \href {https://ui.adsabs.harvard.edu/abs/2007Sci...318.1574D} {318, 1574}

\bibitem[\protect\citeauthoryear{{De Pontieu} et~al.,}{{De Pontieu}
  et~al.}{2007b}]{bart_2007}
{De Pontieu} B.,  et~al., 2007b, \mn@doi [Science] {10.1126/science.1151747},
  \href {https://ui.adsabs.harvard.edu/abs/2007Sci...318.1574D} {318, 1574}

\bibitem[\protect\citeauthoryear{{De Pontieu}, {Hansteen}, {Rouppe van der
  Voort}, {van Noort}  \& {Carlsson}}{{De Pontieu}
  et~al.}{2007c}]{DePontieu_2007a}
{De Pontieu} B.,  {Hansteen} V.~H.,  {Rouppe van der Voort} L.,  {van Noort}
  M.,   {Carlsson} M.,  2007c, \mn@doi [\apj] {10.1086/509070}, \href
  {http://adsabs.harvard.edu/abs/2007ApJ...655..624D} {655, 624}

\bibitem[\protect\citeauthoryear{{De Pontieu}, {Mart{\'\i}nez-Sykora}  \&
  {Chintzoglou}}{{De Pontieu} et~al.}{2017}]{De_pontieu_2017a}
{De Pontieu} B.,  {Mart{\'\i}nez-Sykora} J.,   {Chintzoglou} G.,  2017, \mn@doi
  [\apjl] {10.3847/2041-8213/aa9272}, \href
  {https://ui.adsabs.harvard.edu/abs/2017ApJ...849L...7D} {849, L7}

\bibitem[\protect\citeauthoryear{Hansteen, Pontieu, van~der Voort, van Noort
  \& Carlsson}{Hansteen et~al.}{2006}]{Hansteen_2006}
Hansteen V.~H.,  Pontieu B.~D.,  van~der Voort L.~R.,  van Noort M.,   Carlsson
  M.,  2006, \mn@doi [The Astrophysical Journal] {10.1086/507452}, 647, L73

\bibitem[\protect\citeauthoryear{{Heggland}, {De Pontieu}  \&
  {Hansteen}}{{Heggland} et~al.}{2007}]{Heggland_2007}
{Heggland} L.,  {De Pontieu} B.,   {Hansteen} V.~H.,  2007, \mn@doi [\apj]
  {10.1086/518828}, \href
  {https://ui.adsabs.harvard.edu/abs/2007ApJ...666.1277H} {666, 1277}

\bibitem[\protect\citeauthoryear{{Henriques}, {Kuridze}, {Mathioudakis}  \&
  {Keenan}}{{Henriques} et~al.}{2016}]{Henriques_2016}
{Henriques} V.~M.~J.,  {Kuridze} D.,  {Mathioudakis} M.,   {Keenan} F.~P.,
  2016, \mn@doi [\apj] {10.3847/0004-637X/820/2/124}, \href
  {https://ui.adsabs.harvard.edu/abs/2016ApJ...820..124H} {820, 124}

\bibitem[\protect\citeauthoryear{{Joshi} \& {Rouppe van der Voort}}{{Joshi} \&
  {Rouppe van der Voort}}{2022}]{Joshi_2022}
{Joshi} J.,  {Rouppe van der Voort} L. H.~M.,  2022, arXiv e-prints, \href
  {https://ui.adsabs.harvard.edu/abs/2022arXiv220308172J} {p. arXiv:2203.08172}

\bibitem[\protect\citeauthoryear{{Langangen}, {De Pontieu}, {Carlsson},
  {Hansteen}, {Cauzzi}  \& {Reardon}}{{Langangen}
  et~al.}{2008}]{Langangen_2008}
{Langangen} {\O}.,  {De Pontieu} B.,  {Carlsson} M.,  {Hansteen} V.~H.,
  {Cauzzi} G.,   {Reardon} K.,  2008, \mn@doi [\apjl] {10.1086/589442}, \href
  {https://ui.adsabs.harvard.edu/abs/2008ApJ...679L.167L} {679, L167}

\bibitem[\protect\citeauthoryear{{McIntosh}, {de Pontieu}, {Carlsson},
  {Hansteen}, {Boerner}  \& {Goossens}}{{McIntosh}
  et~al.}{2011}]{McIntosh_2011}
{McIntosh} S.~W.,  {de Pontieu} B.,  {Carlsson} M.,  {Hansteen} V.,  {Boerner}
  P.,   {Goossens} M.,  2011, \mn@doi [\nat] {10.1038/nature10235}, \href
  {https://ui.adsabs.harvard.edu/abs/2011Natur.475..477M} {475, 477}

\bibitem[\protect\citeauthoryear{{Nived} et~al.,}{{Nived}
  et~al.}{2022}]{Nived2022}
{Nived} V.~N.,  et~al., 2022, \mn@doi [\mnras] {10.1093/mnras/stab3277}, \href
  {https://ui.adsabs.harvard.edu/abs/2022MNRAS.509.5523N} {509, 5523}

\bibitem[\protect\citeauthoryear{{Pereira} et~al.,}{{Pereira}
  et~al.}{2014}]{Pereira_2014}
{Pereira} T.~M.~D.,  et~al., 2014, \mn@doi [\apjl]
  {10.1088/2041-8205/792/1/L15}, \href
  {https://ui.adsabs.harvard.edu/abs/2014ApJ...792L..15P} {792, L15}

\bibitem[\protect\citeauthoryear{{Roberts}}{{Roberts}}{1945}]{Roberts_1945}
{Roberts} W.~O.,  1945, \mn@doi [\apj] {10.1086/144699}, \href
  {https://ui.adsabs.harvard.edu/abs/1945ApJ...101..136R} {101, 136}

\bibitem[\protect\citeauthoryear{{Samanta} et~al.,}{{Samanta}
  et~al.}{2019}]{Samanta_2019}
{Samanta} T.,  et~al., 2019, \mn@doi [Science] {10.1126/science.aaw2796}, \href
  {https://ui.adsabs.harvard.edu/abs/2019Sci...366..890S} {366, 890}

\bibitem[\protect\citeauthoryear{{Scharmer}, {Bjelksjo}, {Korhonen}, {Lindberg}
   \& {Petterson}}{{Scharmer} et~al.}{2003}]{Scharmer_2003}
{Scharmer} G.~B.,  {Bjelksjo} K.,  {Korhonen} T.~K.,  {Lindberg} B.,
  {Petterson} B.,  2003, in {Keil} S.~L.,  {Avakyan} S.~V.,  eds,  Society of
  Photo-Optical Instrumentation Engineers (SPIE) Conference Series Vol. 4853,
  Innovative Telescopes and Instrumentation for Solar Astrophysics. pp 341--350

\bibitem[\protect\citeauthoryear{{Scharmer} et~al.,}{{Scharmer}
  et~al.}{2008}]{Scharmer_2008}
{Scharmer} G.~B.,  et~al., 2008, \mn@doi [\apjl] {10.1086/595744}, \href
  {http://adsabs.harvard.edu/abs/2008ApJ...689L..69S} {689, L69}

\bibitem[\protect\citeauthoryear{{Sekse}, {Rouppe van der Voort}  \& {De
  Pontieu}}{{Sekse} et~al.}{2013a}]{sekse_2013a}
{Sekse} D.~H.,  {Rouppe van der Voort} L.,   {De Pontieu} B.,  2013a, \mn@doi
  [\apj] {10.1088/0004-637X/764/2/164}, \href
  {https://ui.adsabs.harvard.edu/abs/2013ApJ...764..164S} {764, 164}

\bibitem[\protect\citeauthoryear{{Sekse}, {Rouppe van der Voort}, {De Pontieu}
  \& {Scullion}}{{Sekse} et~al.}{2013b}]{sekse_2013b}
{Sekse} D.~H.,  {Rouppe van der Voort} L.,  {De Pontieu} B.,   {Scullion} E.,
  2013b, \mn@doi [\apj] {10.1088/0004-637X/769/1/44}, \href
  {https://ui.adsabs.harvard.edu/abs/2013ApJ...769...44S} {769, 44}

\bibitem[\protect\citeauthoryear{{Sekse}, {Rouppe van der Voort}, {De Pontieu}
  \& {Scullion}}{{Sekse} et~al.}{2013c}]{sekse_2013}
{Sekse} D.~H.,  {Rouppe van der Voort} L.,  {De Pontieu} B.,   {Scullion} E.,
  2013c, \mn@doi [\apj] {10.1088/0004-637X/769/1/44}, \href
  {https://ui.adsabs.harvard.edu/abs/2013ApJ...769...44S} {769, 44}

\bibitem[\protect\citeauthoryear{{Shetye}, {Doyle}, {Scullion}, {Nelson},
  {Kuridze}, {Henriques}, {Woeger}  \& {Ray}}{{Shetye}
  et~al.}{2016}]{shetye_2016}
{Shetye} J.,  {Doyle} J.~G.,  {Scullion} E.,  {Nelson} C.~J.,  {Kuridze} D.,
  {Henriques} V.,  {Woeger} F.,   {Ray} T.,  2016, \mn@doi [\aap]
  {10.1051/0004-6361/201527505}, \href
  {https://ui.adsabs.harvard.edu/abs/2016A&A...589A...3S} {589, A3}

\bibitem[\protect\citeauthoryear{{Shetye}, {Verwichte}, {Stangalini}  \&
  {Doyle}}{{Shetye} et~al.}{2021}]{Shetye_2021}
{Shetye} J.,  {Verwichte} E.,  {Stangalini} M.,   {Doyle} J.~G.,  2021, \mn@doi
  [\apj] {10.3847/1538-4357/ac1a12}, \href
  {https://ui.adsabs.harvard.edu/abs/2021ApJ...921...30S} {921, 30}

\bibitem[\protect\citeauthoryear{{Skogsrud}, {Rouppe van der Voort}, {De
  Pontieu}  \& {Pereira}}{{Skogsrud} et~al.}{2015}]{skogsrud_2015}
{Skogsrud} H.,  {Rouppe van der Voort} L.,  {De Pontieu} B.,   {Pereira}
  T.~M.~D.,  2015, \mn@doi [\apj] {10.1088/0004-637X/806/2/170}, \href
  {https://ui.adsabs.harvard.edu/abs/2015ApJ...806..170S} {806, 170}

\bibitem[\protect\citeauthoryear{{Srivastava} et~al.,}{{Srivastava}
  et~al.}{2017}]{srivastava_2017}
{Srivastava} A.~K.,  et~al., 2017, \mn@doi [Scientific Reports]
  {10.1038/srep43147}, \href
  {https://ui.adsabs.harvard.edu/abs/2017NatSR...743147S} {7, 43147}

\bibitem[\protect\citeauthoryear{{Tian} et~al.,}{{Tian}
  et~al.}{2014}]{Tian_2014}
{Tian} H.,  et~al., 2014, \mn@doi [Science] {10.1126/science.1255711}, \href
  {https://ui.adsabs.harvard.edu/abs/2014Sci...346A.315T} {346, 1255711}

\bibitem[\protect\citeauthoryear{{Tsiropoula}, {Tziotziou}, {Kontogiannis},
  {Madjarska}, {Doyle}  \& {Suematsu}}{{Tsiropoula}
  et~al.}{2012}]{Tsiropoula_2012}
{Tsiropoula} G.,  {Tziotziou} K.,  {Kontogiannis} I.,  {Madjarska} M.~S.,
  {Doyle} J.~G.,   {Suematsu} Y.,  2012, \mn@doi [\ssr]
  {10.1007/s11214-012-9920-2}, \href
  {https://ui.adsabs.harvard.edu/abs/2012SSRv..169..181T} {169, 181}

\bibitem[\protect\citeauthoryear{{Vilangot Nhalil}, {Nelson}, {Mathioudakis},
  {Doyle}  \& {Ramsay}}{{Vilangot Nhalil} et~al.}{2020}]{paper_nano_flare}
{Vilangot Nhalil} N.,  {Nelson} C.~J.,  {Mathioudakis} M.,  {Doyle} J.~G.,
  {Ramsay} G.,  2020, \mn@doi [\mnras] {10.1093/mnras/staa2897}, \href
  {https://ui.adsabs.harvard.edu/abs/2020MNRAS.499.1385V} {499, 1385}

\bibitem[\protect\citeauthoryear{{Zaqarashvili} \&
  {Skhirtladze}}{{Zaqarashvili} \& {Skhirtladze}}{2008}]{Zaqarshvili_2008}
{Zaqarashvili} T.~V.,  {Skhirtladze} N.,  2008, \mn@doi [\apjl]
  {10.1086/591524}, \href {http://adsabs.harvard.edu/abs/2008ApJ...683L..91Z}
  {683, L91}

\bibitem[\protect\citeauthoryear{{de Pontieu} et~al.,}{{de Pontieu}
  et~al.}{2007}]{2007PASJ...59S.655D}
{de Pontieu} B.,  et~al., 2007, \mn@doi [\pasj] {10.1093/pasj/59.sp3.S655},
  \href {http://adsabs.harvard.edu/abs/2007PASJ...59S.655D} {59, S655}

\bibitem[\protect\citeauthoryear{{van Noort}, {Rouppe van der Voort}  \&
  {L{\"o}fdahl}}{{van Noort} et~al.}{2005}]{Noort_2005}
{van Noort} M.,  {Rouppe van der Voort} L.,   {L{\"o}fdahl} M.~G.,  2005,
  \mn@doi [\solphys] {10.1007/s11207-005-5782-z}, \href
  {http://adsabs.harvard.edu/abs/2005SoPh..228..191V} {228, 191}

\makeatother
\end{thebibliography}
\end{document}